\documentclass[
    ,final            
  ]
  {aipproc}

\usepackage{amssymb}
\usepackage{amsmath}

\layoutstyle{6x9}

\begin{document}

\title{The Brightest of Reionizing Galaxies (BoRG) survey}


\author{Michele Trenti}{ address={Kavli Fellow, Institute of Astronomy
    and Kavli Institute for Cosmology, University of Cambridge,
    Madingley Road, Cambridge CB3 0HA, United Kingdom} }

\begin{abstract}

  Until now, investigating the early stages of galaxy formation has
  been primarily the realm of theoretical modeling and computer
  simulations, which require many physical ingredients and are
  challenging to test observationally.  However, the latest Hubble
  Space Telescope observations in the near infrared are shedding new
  light on the properties of galaxies within the first billion years
  after the Big Bang, including our recent discovery of the most
  distant proto-cluster of galaxies at redshift $z\sim 8$. Here, I
  compare predictions from models of primordial and metal-enriched
  star formation during the dark ages with the latest Hubble
  observations of galaxies during the epoch of reionization. I focus
  in particular on the luminosity function and on galaxy clustering as
  measured from our Hubble Space Telescope Brightest of Reionizing
  Galaxies (BoRG) survey. BoRG has the largest area coverage to find
  luminous and rare $z\sim8$ sources that are among the first galaxies
  to have formed in the Universe.

\end{abstract}

\maketitle


\section{Introduction}

\begin{figure}
\includegraphics[height=.32\textheight]{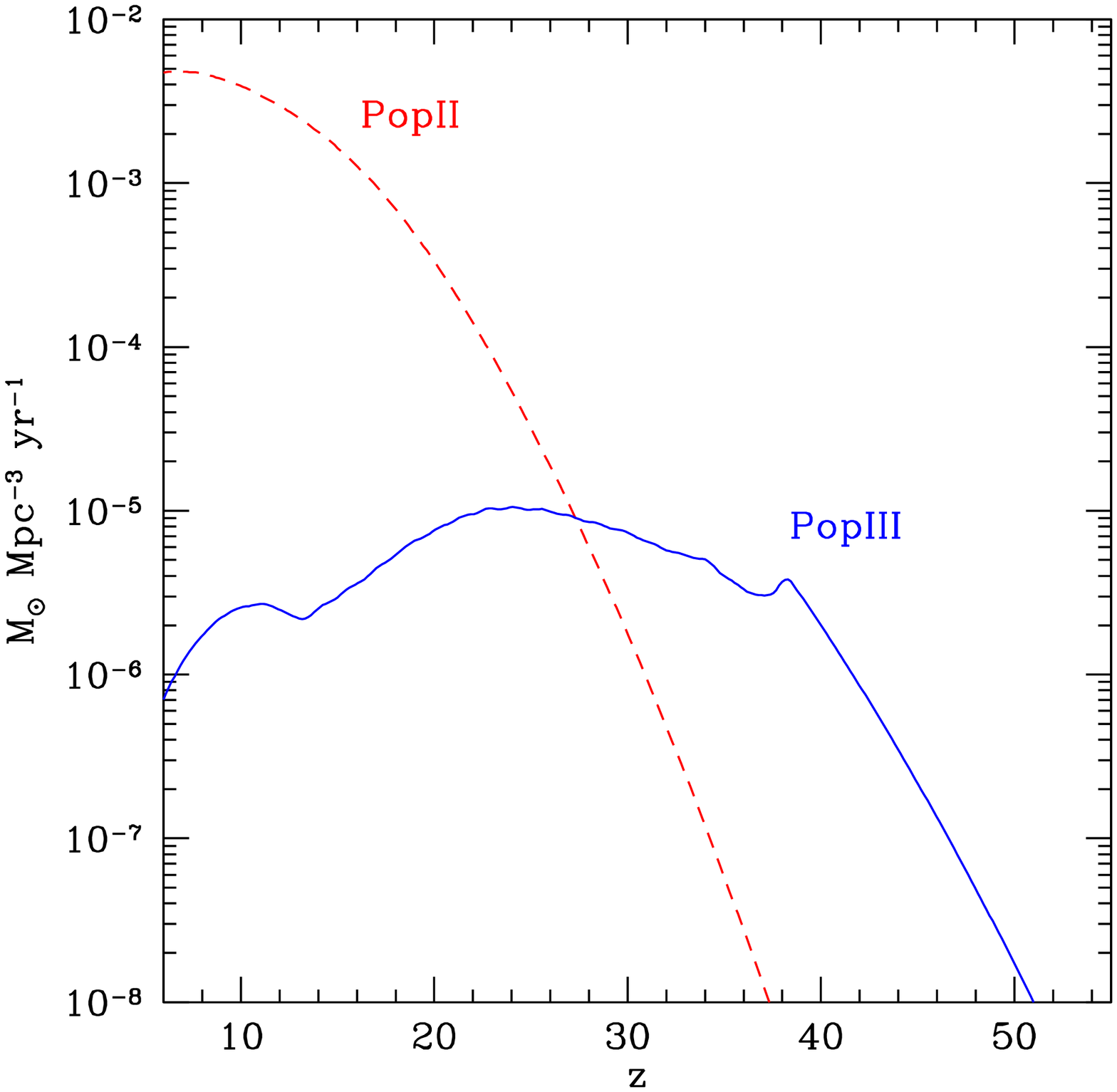}
  \includegraphics[height=.32\textheight]{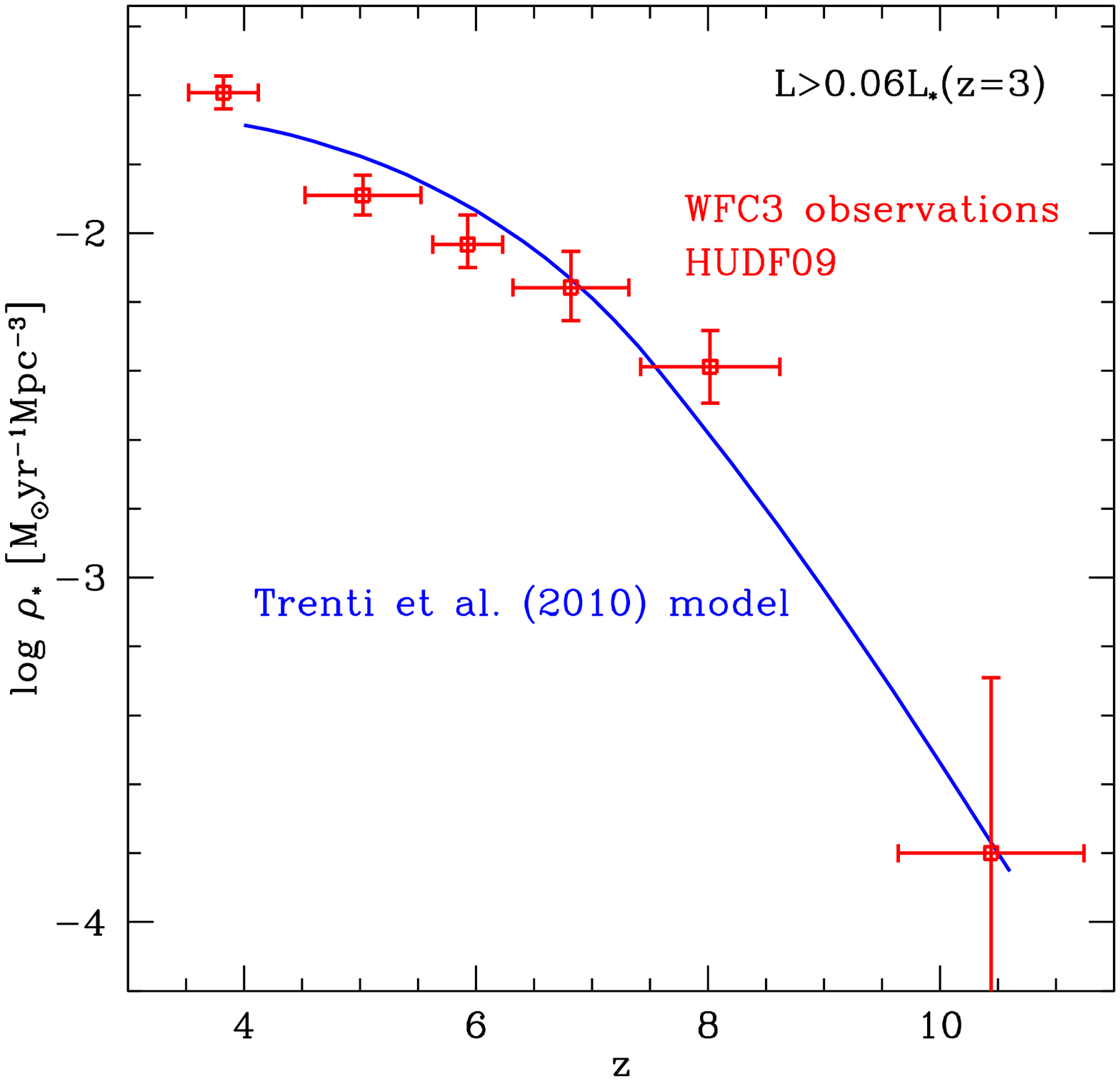}
  \caption{Left panel: Pop III (blue solid line) and Pop II (red
    dashed line) star formation rates based on an analytical model
    including chemical enrichment and radiative feedback (adapted from
    \citep{ts09}). Radiative feedback in the LW bands keeps the Pop
    III star formation rate approximately self-regulated from $z\sim
    35$ to $z\sim 20$. Right panel: Evolution of SFR density at high
    redshift, adapted from WFC3 data of \citep{oesch12}, with
    luminosity density converted to $\dot{\rho}_*$ (see their Table
    3). The solid blue curve shows the predicted evolution from a
    conditional luminosity function model \citep{trenti10} based on
    the evolution of the underlying dark-matter halo mass
    function. The model is successful in reproducing the observed
    $\dot{\rho}_*$ and captures the rapid drop from $z\sim8$ to
    $z\sim10$. }\label{fig:sfr}
\end{figure}

The first billion years after the Big Bang represent a key area of
astrophysics, with interesting problems, open questions and potential
for unexpected and unusual discoveries as highlighted by the 2010
Astronomy Decadal Survey
Report\footnote{\url{http://www.nap.edu/catalog.php?record_id=12951}}. 
Significant progress has been made in the field in the last few
years. Numerical simulations are following formation of the first
stars at progressively high resolution (see reviews by Yoshida and
Abel in this volume) and simulations of first galaxies from first
principles are now growing in dynamic range
\citep{romanodiaz11,wise11}. Theoretical models can be constructed to
investigate and predict the star formation rate during and after the
epoch of Reionization at $z\gtrsim 4$ \citep{ts09,trenti10} (see
Fig.~\ref{fig:sfr}).

Observationally, the quest for direct detection of first stars seems
very difficult, with these sources being too faint (a $100~M_{\odot}$
Pop III star at $z>6$ has observed magnitude $M_{AB}\gtrsim
38$). However, prospects of indirect detection of metal-free stars are
intriguing, either through high-$z$ supernovae, or if gravitational
lensed and living in small clusters \citep{tss09,zack12}.  In terms of
observations of high-redshift galaxies, recent progress has been done
thanks to the installation of Hubble's WFC3, leading to the
identification of large samples of galaxies at $z\sim 7-10$ (see
review by Bouwens in this volume).

In this contribution, I briefly summarize some of my recent results on
star and galaxy formation at high redshift, starting from a short
discussion of metal-free versus metal enriched star formation during
the epoch of reionization (and spatial inhomogeneities in chemical
enrichment), and then focusing on observations of the most overdense
regions in the universe at $z\sim8$ through the Brightest of
Reionizing Galaxies survey.

\section{Star formation during and after the dark ages}

The first stages of star formation at $z\gtrsim 20$ have very well
defined initial conditions, so simple analytical modeling can make
clear predictions of the relative interplay between metal-free and
metal-enriched star formation. With a model for molecular hydrogen
cooling, radiative feedback in the Lyman Werner (LW) bands, plus
chemical enrichment (from metal free to above critical metallicity of
$Z\sim 10^{-3.5} Z_{\odot}$), we have predicted the formation rate of
Pop III (metal-free) and Pop II (metal-enriched) stars at very high
redshift \citep{ts09}. Star formation begins very early in the
Universe (at $z\gtrsim 60$, see also \citep{ts07a}) and the Population
III star formation rate rises exponentially with redshift until
sufficient radiative feedback in the LW bands sets in, at $z\sim 35$
(Fig. 1). Assuming massive Pop III stars, chemical enrichment of the
interstellar medium leads to Pop II star formation to become the
dominant already at $z\sim 25$. Yet, because of the non-linear
relation between redshift and star formation, most Pop III stars are
produced at $z\lesssim 20$.

At lower redshift ($z\lesssim 10$), a simple analytical model for
chemical enrichment is however no longer able to capture the
significant large scale structure that develops in the Universe, with
voids of several Mpc$^3$ (comoving) in volume and the rising
importance of metal outflows in winds. Cosmological simulations are
required to capture the transition from metal-free to metal-enriched
star formation at these later times. This approach shows that Pop III
stars continue forming into low-density regions
\citep{tss09,tornatore07}, while overdense regions, which had earlier
than average Pop III star formation \citep{tss08}, are forming massive
galaxies already at $z\sim 10$. These simulated galaxies have stellar
masses of about $10^9 M_{\odot}$, in the range close to those that can
be observed at slightly later times with the current generation of
Space Telescopes (Hubble and Spitzer, e.g., see \citep{gonzalez11}).

\section{The Brightest of Reionizing Galaxies Survey}

\begin{figure}
  \includegraphics[height=.32\textheight]{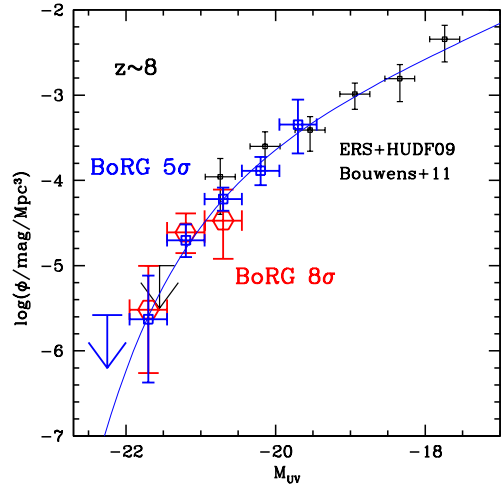}
  \caption{Luminosity Function at $z\sim 8$ from the BoRG survey for
    our $5\sigma$ and $8\sigma$ catalogs (blue and red, respectively),
    combined with the stepwise LF derived by \citep{bouwens11} for
    the ERS+HUDF09 dataset (black points).  The blue line is the
    best-fit Schechter LF from combining the BoRG+ERS+HUDF09 dataset,
    providing the widest dynamic range in luminosity that is currently
    available.  Our results are consistent with a Schechter form of
    the UV LF across all dynamic range probed and the faint-end slope
    is steep: $\alpha =-1.98 \pm 0.23$.}\label{fig:borgLF}
\end{figure}

To investigate galaxy formation during the epoch of reionization in
the most massive dark matter halos, expected to host the brightest
galaxies such as those simulated in \citep{romanodiaz11}, we have
launched a large observational campaign with the Hubble Space
Telescope (HST) to image random pointings of the sky to $m_{AB} \sim
27$ in the near-IR. The Brightest of Reionizing Galaxies Survey is a
four-bands HST survey in the optical (F606W) and near-IR (F098M,
F125W, F160W) which identifies $z\sim 8$ galaxies as F098M-dropout
(Y-dropout) sources using the Lyman break technique
\citep{steidel1996}.

A full description of the survey and of its results is given in
\citep{trenti11,trenti12,bradley12}. To summarize the status to date,
we observed 59 pointings for a total area of $274$ arcmin$^2$, which
is the largest available to search for $z\sim 8$ galaxies (compared to
$\sim 100$ arcmin$^2$ of CANDELS, see \citep{oesch12b}). From this
area we identified 33 $z\sim 8$ galaxy candidates at the bright end of
the luminosity function. Combining the BoRG data with the ultradeep
observations by \citep{bouwens11}, we constructed a global fit of the
galaxy luminosity function at $z\sim 8$ with a Schechter function
($\phi(L) = \frac{\phi_*}{L_*} \left (\frac{L}{L_*} \right)^{\alpha}
\exp{(-L/L_*)}$), shown in Fig.~\ref{fig:borgLF}. The best fit has a
very steep faint-end slope, $\alpha=-1.98\pm0.23$, which suggests that
galaxies fainter than the current detection limit of HST at
$M_{AB}\sim -18$ contribute significantly to the total ionizing flux
emitted at the time (see also the discussion of the same issue in a
theoretical context in \citep{trenti10}).

\section{The Most Distant Galaxy Protocluster}

\begin{figure}
\includegraphics[height=.54\textheight]{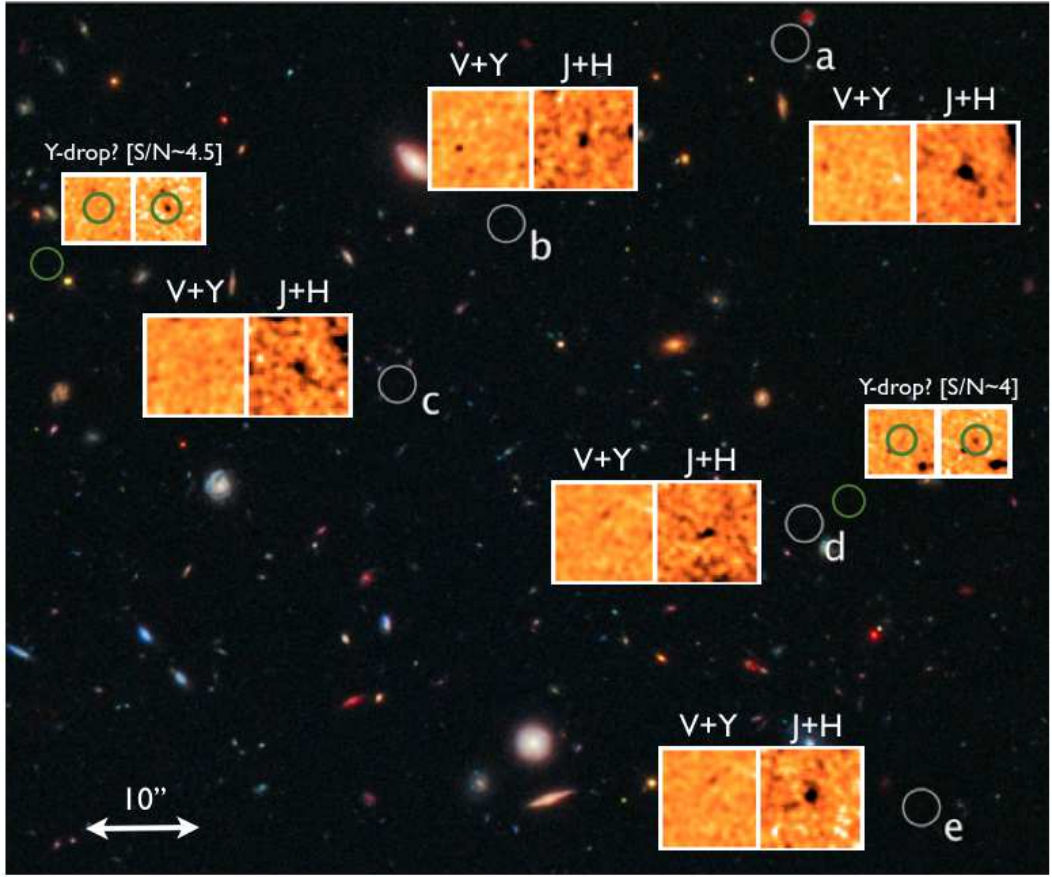}
\caption{VYJH color-composite image of the $z\sim 8$ BoRG58
  protocluster region, with postage stamp images ($3''.2\times3''.2$)
  of the five $z\sim 8$ members of the overdensity (sources a-e)
  discussed in \citep{trenti12}. Also indicated are the locations of
  two additional Y-dropout candidates that are currently just missing
  the $5\sigma$ detection threshold to be included in our
  catalog. Upcoming deeper HST observations (GO-12905, PI Trenti) will
  confirm these additional candidates at $S/N\sim 10$ and are expected
  to discover additional new fainter members of the
  overdensity.}\label{fig:protocluster}
\end{figure}


The BoRG survey is not only counting galaxies at the bright end of the
luminosity function, but it is also identifying special regions in the
Universe, ``cluster construction zones'' that will evolve into the
most massive structures today. The brightest $z\sim 8$ BoRG candidates
are expected to live in highly biased dark matter halos
($M_{DM}\gtrsim 5\times10^{11}~M_{\odot}$, bias $b\sim9$ at $z\sim
8$). As a consequence, they are surrounded by overdensities of fainter
galaxies \citep{munoz08,robertson10}. We verified this prediction at
$>99.9\%$ confidence from BoRG based on field-to-field variations of
$z\sim 8$ number counts and identified a protocluster candidate
\citep{trenti12}. The best $z\sim 8$ source in the BoRG survey, a
$m_{AB}=25.8\pm 0.1$ galaxy more than one magnitude brighter than
$L_*(z=8)$ is surrounded by four additional F098M-dropouts in its
proximity with $S/N>5$, within a region of diameter $d\approx 1'$ (or
$\sim 3$ Mpc comoving) that on average is expected to contain only
$n\sim 0.2$ such galaxies at $z\sim 8$
(Fig.~\ref{fig:protocluster}). With $\langle m_{AB}\rangle=27.1$, the
four sources are $L\sim L_*(z=8)$ galaxies, about $\sim1.3$ magnitudes
fainter than their bright companion and consistent with being at its
redshift. Our detailed statistical analysis, which includes comparison
with mock catalogs from cosmological simulations, demonstrates that
the overdensity is a physical structure of $\sim 3-4$ comoving Mpc in
size at $99.97\%$ confidence. Its discovery was expected based on the
comoving cosmic volume probed by the survey ($V\sim 5\times
10^{5}~\mathrm{Mpc^3}$). The region started forming the first stars at
$z>30$, assembled a significant fraction of its mass by $z\sim15$ and
will evolve by $z=0$ into a massive galaxy cluster with $M>2\times
10^{14}M_{\odot}$ based on our modeling \citep{trenti12}. Hence, these
HST observations are both showing a glimpse of a region that hosted
very early star formation, as well as of the infancy of a future
massive galaxy cluster.

\section{Future prospects}

BoRG is acquiring new data over the course of 2012 through Cycle 19
observations (HST-GO 12572, PI Trenti). The final area of the survey
will grow to about $400$ arcmin$^2$ in the short term, leading to the
discovery of more bright $z\sim 8$ candidates, possibly highly
clustered as in the case of the protocluster field. Recently approved
HST observations of the BoRG protocluster (HST-GO 12905) are expected
to lead to the discovery of $6-10$ new members of the overdensity,
confirming theoretical predictions that clustering extends to smaller
mass halos containing fainter galaxies \citep{trenti12}.  Ground-based
spectroscopy (e.g. MOSFIRE at Keck) has the potential to provide the
first solid detections of Ly$\alpha$ emission from sources at $z\sim
8$. Doing so would confirm the redshift of photometric $z\sim 8$
candidates such as those identified by BoRG. More importantly this
would also constrain models for the evolution of the neutral hydrogen
fraction, which affects the Equivalent Width distribution of
Ly$\alpha$ emission \citep{treu12}. Finally, Spitzer IRAC observations
of luminous galaxies in high-$z$ overdensities have the potential to
unveil the properties of stellar populations of some of the first
galaxies formed in the Universe, given the biased star formation
history of these regions. All this array of future observations,
combined with theoretical and numerical modeling, will pave the way to
extending the study of galaxy formation to $z\sim 15$ once the {\it
  James Webb Space Telescope} will be launched later this Decade.


\begin{theacknowledgments}
  It is a pleasure to thank the Scientific and Local Organizing
  Committees, and especially the LOC chair Kazu Omukai, for the
  excellent organization of a productive and interesting meeting.
\end{theacknowledgments}



\bibliographystyle{aipproc}   


\end{document}